\documentclass[slac_one]{revtex4}
\usepackage{graphicx}
\usepackage{fancyhdr}
\begin{document}

\title{The Origin of Mass in QCD\footnote{MSUHEP-041115}}

%

\author{R. Sekhar Chivukula\footnote{e-mail: sekhar@msu.edu}}
\affiliation{Department of Physics and Astronomy, Michigan State University,
East Lansing, MI 48824-2320, USA}

\begin{abstract}
In this talk I discuss the origin of mass in quantum chromodynamics in the context of
the classical and quantum symmetries of the theory.

\end{abstract}

\maketitle

\thispagestyle{fancy}


\section{Introduction}

\begin{quote}

``As we know, \\
There are known knowns. \\
There are things we know we know. \\
We also know \\
There are known unknowns. \\
That is to say \\
We know there are some things \\
We do not know. \\
But there are also unknown unknowns, \\
The ones we don't know \\
We don't know." \\
\emph{--  Donald Rumsfeld, U.S. Secretary of Defense, Feb. 12, 2002}

\end{quote}


Using the classification suggested by Donald Rumsfeld, the subject of
the majority of this conference is the ``known unknowns" -- those questions which
can usefully be framed in the context of the standard model of particle physics
(and cosmology), but whose answers remain elusive. The unknown, unknowns are
the subject of philosophy. 

In contrast the subject of this talk, quantum
chromodynamics or QCD,  is a ``known known." Why should we spend time
studying this topic at this conference? I hope to convince you in
the course of this lecture that there are at least three reasons to do so: 

\begin{enumerate}

\item  The elucidation of the strong force is one of the great intellectual triumphs
of quantum field theory \cite{nobel}.

\item QCD is the only experimentally studied strongly-interacting quantum field theory
and, as such, illustrates many subtle issues in field theory (many of which are the
subject of this lecture).

\item QCD is a paradigm for the sort of strongly-interacting field theories which may be
involved in the solution of the ``known unknowns" discussed in the rest of this conference.

\end{enumerate} 

After reviewing the basics of QCD, the bulk of the lecture will discuss the origin of mass
in QCD in terms of the classical symmetries of the QCD Lagrangian and their quantum 
analogs, and I will conclude with some applications of the properties of QCD (or QCD-like
theories) to other issues in particle physics.

\subsection{What is Mass?}

There are many overlapping definitions of mass, arising from
\begin{itemize}

\item Newton's Second Law: $\vec{F} = m \vec{a}$.

\item The Relativistic Dispersion Relation: $E^2=p^2 + m^2$.

\item Newton's Principle of Equivalence: $m_{grav} = m$.

\item Einstein's Principle of Equivalence: $G_{\mu\nu} \propto T_{\mu \nu}$.

\end{itemize}
Each of these definitions is subtle and interesting, and subject to a range of 
important experimental tests and theoretical limitations. In this talk, we will be
interested in how the theory of the strong interations -- quantum chromodynamics --
affects the masses of the physical particles as inferred by {\it any} of these definitions.
Even more interesting: each of the definitions above involve kinematic
tests on individual particles -- but the strong constituents of matter, quarks and
gluons, are confined! What, precisely, do we mean by the mass of these
particles?

\subsection{What is QCD?}

\begin{figure*}[t]
\centering
\includegraphics[width=50mm]{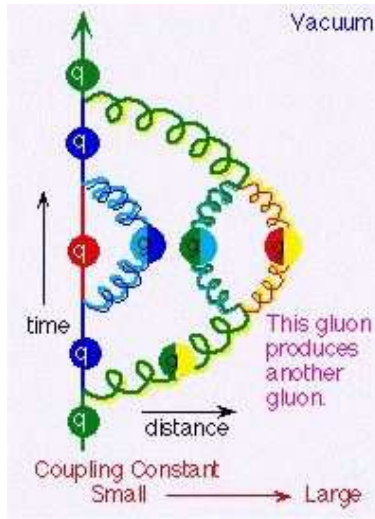}
\caption{Contribution to ``self-energy'' of the quark, illustrating the coupling of gluons
to themselves.} \label{gluonselfcoupling}
\end{figure*}

\begin{figure*}[t]
\centering
\includegraphics[width=135mm]{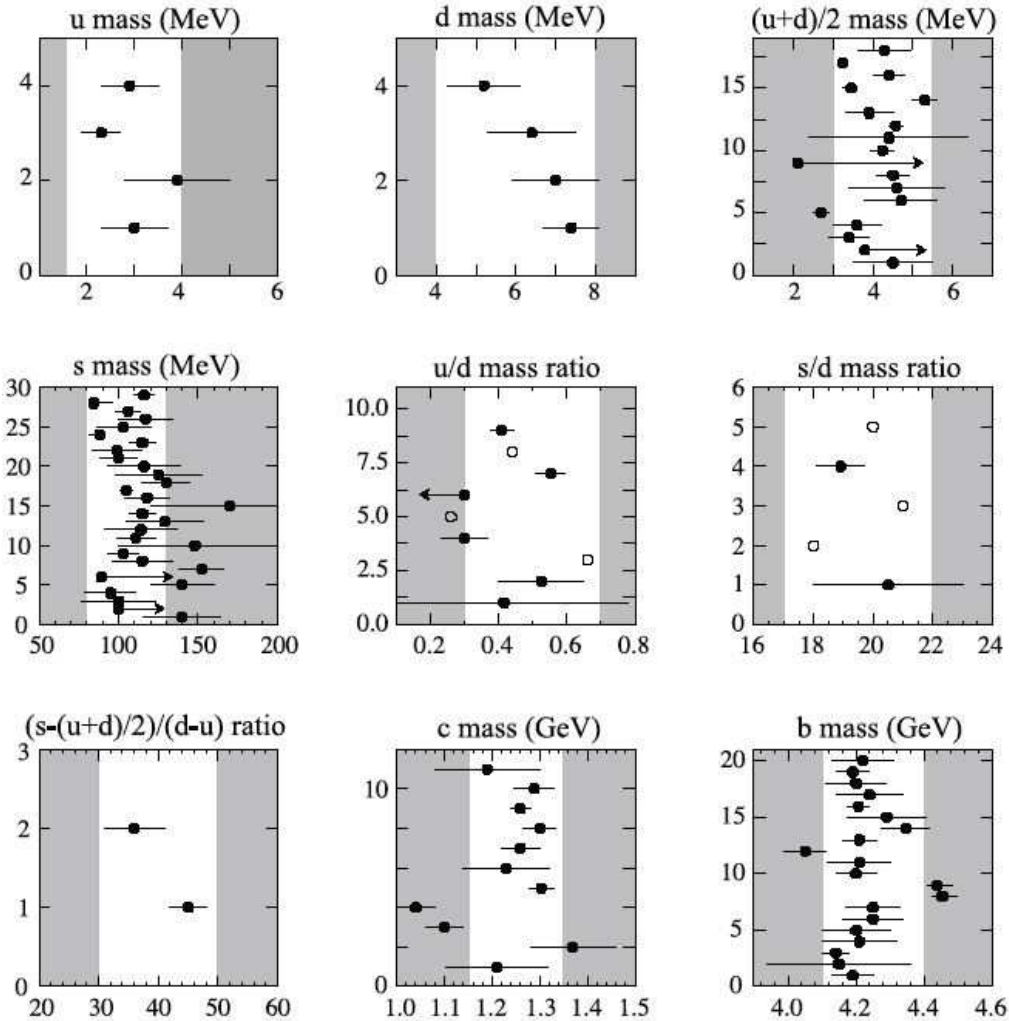}
\caption{Summary of Lagrangian quark masses from Manohar and Sachrajda in
\protect\cite{Eidelman:2004wy}.} \label{quarkmasses}
\end{figure*}

Quantum chromodynamics is the $SU(3)$ Yang-Mills theory of interacting quarks
and gluons, and may be summarized by the Lagrangian
\begin{equation}
{\cal L}_{QCD} = -\,{1\over 4} F^{(a)}_{\mu\nu} F^{(a)\mu\nu} 
+ i\sum_q \bar{\psi}_{qi} \left[ \gamma^\mu  
(D_\mu)^i_j  - m_q \delta^i_j \right] \psi^j_{q}~,
\label{eq:qcdlagrangian}
\end{equation}
where the $\psi^j_{q}$ are the quark fields (labeled by flavor $q$ and color $j$) transforming in the fundamental ($3$) representation of the color $SU(3)$ gauge group and $m_q$ is the 
``Lagrangian'' mass of quark $q$.  $D_\mu$
is the covariant derivative
\begin{equation}
(D_\mu)^i_j = \delta^i_j \partial_\mu + i g_s \sum_a {(\lambda^a)^i_j \over 2} A^a_\mu~,
\end{equation}
where $\lambda^a$ ($a=1,\ldots, 8$) are the $SU(3)$ Gell-Mann matrices, the $A^a_\mu$ are the
gluon fields,  and $g_s$ is the QCD (strong) coupling constant. Finally, $F^{(a)}_{\mu\nu}$ is
the gluon field-strength tensor
\begin{equation}
F^{(a)}_{\mu\nu} = \partial_\mu A^a_\nu - \partial_\nu A^a_\mu - g_s f_{abc} A^b_\mu A^c_\nu~,
\label{eq:fieldstrength}
\end{equation}
where the constants $f_{abc}$ are the $SU(3)$ structure constants.
A summary of the quarks and their quantum numbers is given in fig. \ref{quarks}, and a summary
of our knowledge of the Lagrangian quark masses is given in fig. \ref{quarkmasses}.

The distinguishing feature of non-abelian gauge theories like QCD is that the 
gauge-bosons, the gluons, carry charge and couple to themselves. The relevant couplings
arise from the non-linear terms in the field-strength (eqn. \ref{eq:fieldstrength}), and are
illustrated in fig. \ref{gluonselfcoupling}. It is this property of QCD which gives rise to the
numerous non-trivial features of the theory.

\begin{figure*}[t]
\centering
\includegraphics[width=135mm]{quarks.epsf}
\caption{The Quarks.} \label{quarks}
\end{figure*}

\subsection{The Quark Model}

\begin{figure*}[t]
\centering
\includegraphics[width=50mm]{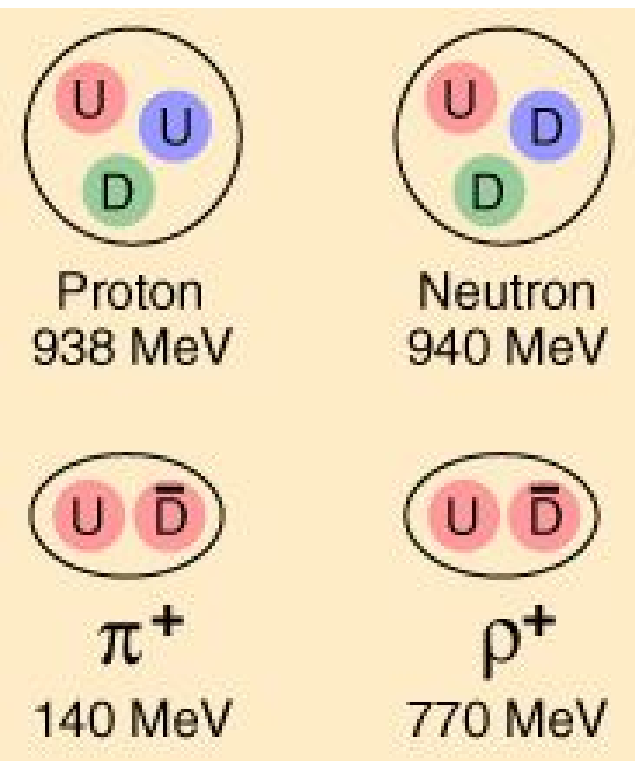}
\caption{Graphical representation of the quark model.} \label{quarkmodel}
\end{figure*}

Explaining the Lagrangian quark masses of fig. \ref{quarkmasses} is the content of the ``flavor
problem'' -- the subject of many talks at this summer school. The Lagrangian masses of the light
quarks, a few to 10  MeV for the up and down quarks, and around 100 MeV for
the strange quark, should be contrasted with what we might expect from the quark
model -- illustrated in fig. \ref{quarkmodel}. Naively, looking at the baryons or heavy mesons, we
expect the quark masses to be of order a third the proton mass or about 300 MeV. While this would seem
reasonable for the baryons or heavy mesons, the quark model doesn't explain why the pions
are so light! So, in so far as the quark masses are concerned, we have three mysteries: 

\begin{itemize}

\item How do we
interpret the Lagrangian masses (or equivalently, what measurements lead to the results in 
fig. \ref{quarkmasses})?

\item How do we interpret the quark model masses of about 300 MeV for the quarks?

\item How do we account for the anomalously light pions?

\end{itemize}

\section{Classical Symmetries of QCD}

Understanding mass in QCD will hinge on understanding of symmetries
of QCD.

\subsection{Space-time Symmetries}

First, we consider the space-time symmetries of QCD:

\begin{itemize}

\item Poincare symmetry: As with all relevant quantum field theories, QCD respects
relativistic invariance -- both Lorentz invariance and translational invariance.

\item As written, the theory respects charge-conjugation, parity, and time-reversal 
invariance\footnote{Stay tuned, however, there is potentially another interaction which
could have been written in eqn. (\protect\ref{eq:qcdlagrangian}) that could affect this
conclusion -- see eqn. (\protect\ref{eq:cpnot}).}.

\item (Approximate) Scale Invariance: Consider the scale transformations
\begin{equation}
x^\mu \rightarrow \lambda x^\mu~,\quad
\psi_q(x) \rightarrow \lambda^{3/2} \psi_q(\lambda x)~, \quad
A^a_\mu(x) \rightarrow \lambda A^a_\mu(\lambda x)~.
\label{eq:scale}
\end{equation}
To the extent that the quark masses are small,\footnote{In what follows, we will
see to what extent the $u$, $d$, and $s$-quark masses are small.} classical
QCD is approximately scale-invariant.

\end{itemize}

\subsection{Global Quark Flavor Symmetries}

Next, consider the global quark symmetries of the theory:

\begin{itemize}

\item Baryon number:
\begin{equation}
\psi_q \rightarrow e^{i\alpha} \psi_q~,
\end{equation}
is an exact symmetry of QCD.

\end{itemize}

To the extent that the $u$, $d$, and $s$ quarks may be considered
equal, we have

\begin{itemize}

\item Approximate $SU(3)_V$ symmetry (Gell-Mann)
\begin{equation}
\left(
\begin{array}{c}
u \cr
d \cr
s
\end{array}
\right) \rightarrow
U
\left(
\begin{array}{c}
u \cr
d \cr
s
\end{array}
\right)~.
\label{eq:vectorsymmetry}
\end{equation}

\end{itemize}

And to the extent the $u$, $d$, and $s$ quarks are light, we have the chiral symmetries

\begin{itemize}

\item Approximate Chiral $SU(3)_L \times SU(3)_R$

\end{itemize}
\begin{equation}
\left(
\begin{array}{c}
u^{L,R} \cr
d ^{L,R}\cr
s^{L,R}
\end{array}
\right) \rightarrow
U_{L,R}
\left(
\begin{array}{c}
u^{L,R} \cr
d^{L,R} \cr
s^{L,R}
\end{array}
\right)~,
\label{eq:chiralsymmetry}
\end{equation}
is an invariance of the quark kinetic energy terms
\begin{equation}
\bar{\psi}_q i \slash\! \! \! \!D \psi_q = 
\bar{\psi}^L_q i \slash\! \! \! \!D \psi^L_q + \bar{\psi}^R_q i \slash\! \! \! \!D \psi^R_q~,
\end{equation}
but not the quark mass terms.
\begin{equation}
m_q \bar{\psi}_q \psi_q \equiv m_q \bar{\psi}^L_q \psi^R_q + m_q \bar{\psi}^R_q \psi^L_q~.
\end{equation}
$SU(3)_V$ is the subgroup of $SU(3)_L \times SU(3)_R$ with $U_L = U_R$. The orthogonal
set of axial transformations -- with $U_L = U^\dagger_R$ -- are often denoted ``$SU(3)_A$", even 
they do not form a group.

\begin{itemize}

\item And, finally, $U(1)_A$
\begin{equation}
\psi^L_q \rightarrow e^{i\alpha} \psi^L_q~,\qquad
\psi^R_q \rightarrow e^{-i\alpha} \psi^R_q~.
\label{eq:u1asymmetry}
\end{equation}

\end{itemize}

\section{No Quantum Scale Invariance}

\subsection{The Running Coupling}

\begin{figure*}[t]
\centering
\includegraphics[width=60mm]{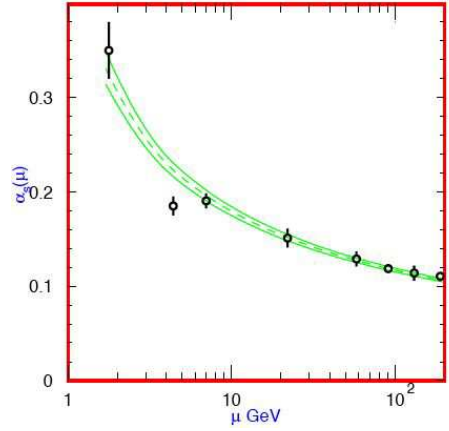}
\caption{Experimental and theoretical values for the running value
of the strong coupling constant \protect\cite{Eidelman:2004wy}.} \label{runningcoupling}
\end{figure*}

One of the fundamental differences between the classical and quantum 
systems is the nature of the vacuum. In quantum field theory, the vacuum is
a polarizable medium. Therefore, the effective ``charge'' measured for any coupling
constant is a function of the scale at which the measurement is made. The
variation of the effective charge as a function of scale is summarized by
the $\beta$-function of the theory. In QCD, the $\beta$-function for the QCD
coupling constant at three-loops (in the $\overline{\rm MS}$ scheme) is given by
\begin{equation}
\mu{\partial \alpha_s \over \partial \mu} = 2 \beta(\alpha_s) =
-{\beta_0\over 2\pi} \alpha^2_s -{\beta_1 \over 4\pi^2} \alpha^3_s
-{\beta_2 \over 64\pi^3} \alpha^4_s - \ldots
\label{eq:betafunction}
\end{equation}
where
\begin{equation}
\beta_0 = 11-{2\over 3}n_f~, \qquad \beta_1 = 51-{19\over 3} n_f~,
\qquad \beta_2 = 2857-{5033\over 9} n_f +{325\over 27}n^2_f~.
\label{eq:betacoefficients}
\end{equation}

Note the negative sign of the QCD $\beta$-function -- this sign is a crucial difference between
non-abelian and abelian gauge-theories: the negative sign can be traced to the contribution
from the self-interactions of the gluons. This results in the behavior of the coupling illustrated
in fig. \ref{runningcoupling}. At higher energies, shorter distances, this results in 
the effective coupling becoming weaker -- asymptotic freedom \cite{nobel} --
which ultimately justifies the parton model description of hadrons at high energies.
In the opposite limit, as one scales to lower energies or larger distances, the effective
coupling grows -- this is sometimes called infrared slavery, and allows for confinement
of color charges. 

\subsection{What is the value of $\bf \alpha_s$? Which Quarks are Light?}

\begin{figure*}[t]
\centering
\includegraphics[width=60mm]{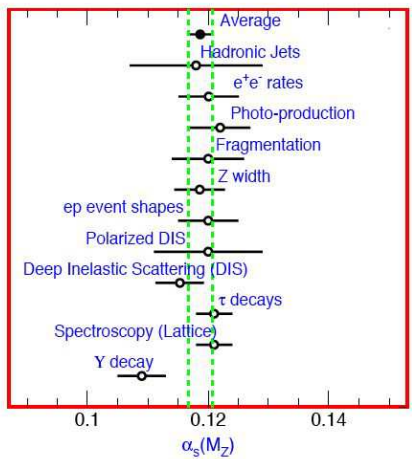}
\caption{The values of $\alpha_s(M_Z)$ and $\Lambda_{QCD}$
as extracted from various experiments \protect\cite{Eidelman:2004wy}.} \label{strongcoupling}
\end{figure*}

Given the value of the strong coupling at one scale, the renormalization group equation
allows for the prediction of its value at any other scale. Conversely, to allow for
the comparison of $\alpha_s$ extracted from different experiments, 
one may use the renormalization group equation
to quote the results of any experiment in terms of $\alpha_s(\mu=M_Z)$.
The solution
of the renormaliztion group equations (at three-loop order), may be written
\begin{eqnarray}
\alpha_s(\mu) &=& {4\pi \over \beta_0 \log(\mu^2/\Lambda_{QCD}^2)}
\left[1-{2\beta_1\over \beta^2_0}\,{\log\left[\log(\mu^2/\Lambda_{QCD}^2)\right] \over 
\log(\mu^2/\Lambda_{QCD}^2)}  +{4\beta^2_1 \over \beta^4_0 \log^2(\mu^2/\Lambda_{QCD}^2)}\right. \\
& \times & \left.\left(\left(\log\left[\log(\mu^2/\Lambda^2_{QCD})\right]-{1\over 2}\right)^2
+{\beta_2 \beta_0\over 8\beta^2_1}-{5\over 4}\right)\right] \nonumber~,
\label{eq:solution}
\end{eqnarray}
and alternatively,  instead of quoting a value of $\alpha_s$ one may quote
the value of a {\it dimensionful} quantity, $\Lambda_{QCD}$.

It is important to realize that a {\it coupling constant} per se is not a directly observable quantity --
only the results of a potential experiment is an observable. The program of physics is to use
the measurements of some finite number of experiments to calculate the results
of others -- coupling constants\footnote{As well as other quantities such as Lagrangian
quark masses which, due to confinement, do not correspond the pole masses of observable
particles.} are simply useful numerical intermediate steps  in the calculations. 
As such, the extracted values of $\alpha_s$ and $\Lambda_{QCD}$
 depend on the calculational scheme chosen  ({\it e.g.}, the results shown 
in fig. \ref{strongcoupling} correspond to the $\overline{\rm MS}$ renormalization prescription).

In particular in relating the value of $\alpha_s(M_Z)$ to $\Lambda_{QCD}$ one must specify
the number of active quark flavors ($n_f$) in eqns. (\ref{eq:betafunction}) and 
(\ref{eq:betacoefficients}). At a scale of order $M_Z$, there are five active flavors
(corresponding to the $u$, $d$, $s$, $c$, and $b$ quarks), and it is therefore conventional
to quote the corresponding value -- this is found \cite{Eidelman:2004wy} to be
\begin{equation}
\Lambda^{(5)}_{\overline{\rm MS}} = 217^{+25}_{-23}\, {\rm MeV}~.
\end{equation}
At low-energies, of order the masses of the lightest baryons, the number of
active flavors is only three (for the $u$, $d$, and $s$), and the corresponding
value is then\footnote{One determines the value of $\Lambda^{(3)}$ from $\Lambda^{(5)}$
by imposing continuity of the coupling constant at the scales of order the heavy quark masses
that are being ``integrated out."}
\begin{equation}
\Lambda^{(3)} \simeq 350\, {\rm MeV}~.
\end{equation}
Examining eqn. (\ref{eq:solution}) and fig. \ref{runningcoupling}, 
we see that this value sets the scale at which
the strong coupling becomes large, and therefore sets the energy scale
at which nonpertubative effects become important.

In our discussion of the symmetries of QCD, we saw that the chiral symmetries
were only approximate and were broken explicitly by Lagrangian quark masses.
These symmetries are useful only to the extent that the symmetry breaking masses
are ``small'' -- in particular, only if the Lagrangian quark masses are small compared
to the  low-energy QCD scale of order 350 MeV. It is in this sense that the $u$,
$d$, and $s$ quarks are light -- and why we don't consider\footnote{It is important to note,
however, that dimensional transmutation is essential to our ability to consider
the heavy quark limit, $\Lambda_{QCD}/m_q \to 0$, which results in symmetries relating
the properties of various $c$ and $b$ mesons and baryons \cite{Isgur:1989vq,Shifman:1986sm,Manohar}.}
the chiral properties
of the $c$, $b$, or $t$ quarks!

\subsection{The Death of Scale Invariance and the Mass-Gap}

The appearance of a dimensional scale related to the dimensionless
coupling of QCD is an example of ``dimensional transmutation'' -- a general
property of quantum field theory.\footnote{For an excellent review of this and
other topics in field theory, see \protect\cite{coleman}.} 

\begin{figure*}[t]
\centering
\includegraphics[width=60mm]{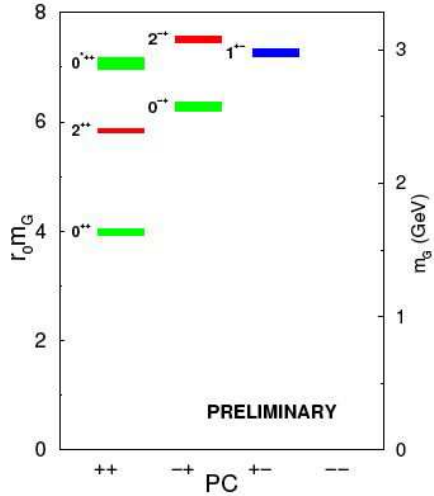}
\caption{Glueball spectrum in QCD without quarks,
from lattice gauge theory \protect\cite{Morningstar:2003ew}.} \label{glueball}
\end{figure*}

Dimensional transmutation
would arise even if there were no quarks -- in this case, the spectrum of the theory
would consist entirely of massive bound states of gluons called glueballs.
In the real world, of course, we cannot eliminate the quarks and experimentally
verify this belief. We can, however, do numerical simulations of such a theory
using lattice gauge theory. The results of such a calculation \cite{Morningstar:2003ew}
are shown in fig. \ref{glueball}.  We see that an $SU(3)$ Yang-Mills theory (without
fermions)  has a ``mass gap'' -- that is, it has no massless excitations.

The existence of a mass gap in quarkless QCD, a theory with no dimensional
parameters in the classical Lagrangian, is a sign that the classical scale invariance
of the theory is broken. More generally, the fact that the QCD coupling runs (fig. \ref{runningcoupling})
and that one can therefore characterize the coupling in terms of 
$\Lambda_{QCD}$, shows that scale invariance in QCD is broken in the quantum theory. 

From Noether's theorem,
we know that any continuous transformation defines an associated current -- and we
know that if this transformation is a {\it symmetry}, the corresponding current is
conserved. We can calculate the current associated with scale transformations,
eqn. (\ref{eq:scale}), $s^\mu$ -- and computing in the quantum theory we find
\begin{equation}
\partial_\mu s^\mu = -\,{\beta\over 2g_s} F^{(a)}_{\mu\nu} F^{(a)\mu\nu} +
\sum_q m_q \bar{\psi}_q \psi_q \neq 0
\end{equation}
The second term in the equation shows what we expect -- the Lagrangian quark
masses explicitly break the scale symmetry. The first term, however, is unexpected --
it shows that the scale invariance is also broken by the fact that the $\beta$-function
of QCD is not zero. The $\beta$-function is an intrinsically quantum effect, and
this result illustrates that scale symmetry is {\it anomalous} -- it is an approximate
symmetry of the classical theory which is explicitly broken by quantum fluctuations!

\section{Chiral Symmetries are broken!}

\subsection{Why the Pions are Light ... }

Unlike scale symmetry, the nonabelian $SU(3)_L \times SU(3)_R$ chiral symmetries
({\it c.f.} eqn. (\ref{eq:chiralsymmetry}))
are good quantum symmetries of a theory with massless quarks -- and these symmetries
are therefore approximate symmetries of the world to the extent that the $u$,
$d$, and $s$ quarks are light. However, the strong low-energy QCD dynamics
rearranges the vacuum and the attractive interactions in the color-singlet
spin-zero channel cause a Bose-Einstein condensate of the quark fields
\begin{equation}
\langle \bar{u}_L u_R \rangle = \langle \bar{d}_L d_R \rangle \approx
\langle \bar{s}_L s_R \rangle \propto \Lambda^3_{QCD} \neq 0~,
\end{equation}
with, because of parity symmetry,  an equal condensate 
for the opposite chirality $RL$ combinations of these fields. The
condensate breaks the individual $SU(3)_{L,R}$ symmetries down
to the vectorial symmetry $SU(3)_V$ ({\it c.f.} eqn. (\ref{eq:vectorsymmetry})).

Chiral symmetry breaking in QCD is an example in which the nonperturbative
quantum dynamics of the theory drives the {\it spontaneous} breaking of
a symmetry. In the three-quark theory, the chiral condensate breaks eight
linearly independent continuous symmetries, and eight corresponding
currents $j^i_{A\mu}$. Goldstone's theorem \cite{Goldstone:1961eq} tells
us that there will be eight low-energy Goldstone bosons ($\pi^i$) associated with these
currents, and we may write
\begin{equation}
j^i_{A\mu} = -f_\pi \partial_\mu \pi^i + \ldots
\end{equation}
where $f_\pi$ is the pion decay constant, approximately 93 MeV in the
normalization chosen here (and the dots correspond to terms with more fields
whose form is determined by symmetry).

In a theory with massless quarks,
$\partial^\mu j^i_{A\mu} \equiv 0$ and the corresponding Goldstone 
bosons would be
massless. 
As the $u$, $d$, and $s$ quarks are light but not massless, we expect
the corresponding particles to be {\it light}.
The lightest strongly-interacting particles are the pions, and
identifying them as the ``would-be'' Goldstone bosons of QCD, 
we explain why they are anomalously light. Treating the quark
masses as perturbations, we find
\begin{eqnarray}
m^2_\pi \propto (m_u + m_d) \Lambda_{QCD} \nonumber \\
m^2_K \propto (m_s + m_{u,d})\Lambda_{QCD} \\
m^2_{\eta} \propto {1\over 3} (m_u + m_d + 4 m_s) \Lambda_{QCD}~,
\end{eqnarray}
where the masses in this expression are to be interpreted as Lagrangian
quark masses (as these are what explicitly break the chiral symmetries) normalized
at energies of order a GeV (indeed, these expressions and their higher-order
relatives are input into fig. \ref{quarkmasses}).

In the limit $m_u=m_d$ there is one relation amongst these three
mass squareds -- this is the Gell-Mann--Okubo relation, and it is well
satisfied for the squareds masses of the pions. Furthermore, 
having identified the pions as Goldstone bosons of chiral symmetry
breaking, the chiral symmetry algebra
implies many relations among pion amplitudes -- so-called current algebra
relations -- which are known to be satisfied within approximately 20\%.

\subsection{... and the Baryons Heavy}

Chiral symmetry breaking also gives us a picture of how to understand
the quark model illustrated in fig. \ref{quarkmodel}. Aside from the pions,
which are light by virtue of being approximate Goldstone bosons, the
hadrons may be thought of as composed of ``constituent quarks'' --
quarks dressed by their interactions with the chiral symmetry breaking QCD
vacuum.  From this point of view, the pions are just different -- they are intrinsically relativistic
approximate Goldstone boson bound states, and are not usefully characterized by
the non-relativistic quark model.

The mass scale associated with the interactions between the quarks and
the vacuum is set by
the (three-quark) value of $\Lambda_{QCD}$, and
the approximate 300 MeV masses of quarks in the quark model should
be interpreted as constituent (dressed) quark masses. 
Note that the constituent quark masses have nothing,
{\it a priori}, to do with the Lagrangian quark masses discussed previously.
Since the Lagrangian quark masses for the $u$ and $d$ quarks are only
of order 10 MeV or less, we see that 99\% of the mass of the proton (and therefore
essentially all the mass of ordinary matter in the universe) arises
\footnote{There
is a potential subtlety in this argument: to the extent that the strange quark is
heavy, the proton and neutron masses could be thought of as arising from a 
``two-flavor'' value of $\Lambda_{QCD}$. Since all representations of $\Lambda_{QCD}$
must predict the same value of $\alpha_s(M_Z)$, this two-quark value 
$\Lambda_{QCD}$ (and therefore the proton mass)
depends indirectly on the value of the strange quark mass. Using
current algebra, this dependence can be related to various pion-nucleon scattering
amplitudes --- this is related to  the so-called ``sigma term" -- and the inferred value
is surprisingly high \protect\cite{Gasser:1980sb}
\begin{equation}
m_s {dm_p \over dm_s} \simeq {\cal O}\left( 10\% \right)~.
\end{equation}
}
from QCD!

\subsection{What about $U(1)_A$?}

The $U(1)_A$ symmetry of eqn. (\ref{eq:u1asymmetry}) would also be broken
by the chiral condensate. If this were truly a symmetry, one would expect
a ninth approximate Goldstone boson. This boson would be an isosinglet pseudoscalar boson.
The lightest candidate is the $\eta'$ which has a mass 958 MeV -- and is not
particularly light! In fact, if $U(1)_A$ is truly a symmetry
of QCD, Weinberg \cite{Weinberg:1975ui}
showed that the mass of the corresponding
approximate Goldstone boson is bounded by $\sqrt{3} m_\pi \simeq 225$ MeV. The
absence of such a ninth approximate Goldstone boson was known as the
$U(1)$-problem.

In fact, $U(1)_A$ is anomalous. Like scale invariance, this classical symmetry
is violated in the quantum theory. As shown by Adler, and Bell and Jackiw
(originally in the context of QED) the divergence of the corresponding current
is not zero \cite{Adler:1969gk,Bell:1969ts}. In the case of the $U(1)_A$ quark current in QCD, one finds
in the massless quark limit
\begin{equation}
\partial_\mu j^\mu_A \propto {g^2_s \over 16\pi^2} \epsilon^{\mu\nu\alpha\beta}
F^{(a)}_{\mu\nu} F^{(a)}_{\alpha\beta}~.
\label{eq:anomaly}
\end{equation}
The axial anomaly was well-known prior to Weinberg's work \cite{Weinberg:1975ui},
however the $F \tilde{F}$ combination of field-strength tensors appearing in
eqn. (\ref{eq:anomaly}) can be shown to be a {\it total derivative}! Therefore, such
an interaction cannot have any effect to any {\it finite} order in perturbation theory.
In principle, one could redefine the current $j^\mu_A$ such that it was conserved
and, as such, it was hard to see why a ninth Goldstone boson was absent.

Fortunately, shortly thereafter,  't Hooft demonstrated \cite{'tHooft:1976fv} that there were nonperturbative
contributions  -- instantons -- which resolved
this problem. 't Hooft showed that, in the semiclassical approximation, there were field
 configurations\footnote{In Euclidean space, but that is a technicality.} of finite action which had the property that
the the integral of $F\tilde{F}$ doesn't vanish. While the semiclassical approximation breaks down
at low-energies in QCD, because the strong coupling becomes large, instantons demonstrate explicitly
that $j^\mu_A$ is broken by nonperturbative quantum effects. $U(1)_A$ is therefore
not a symmetry of QCD, and there is no $U(1)$ problem.

No good deed goes unpunished, however. Having shown that the operator $F\tilde{F}$
could have an effect in QCD (or in any other non-abelian gauge theory, in fact), it is possible
to entertain an additional term in the Lagrangian ${\cal L}_{QCD}$
\begin{equation}
{\cal L}_{CP} = -{g^2_s\,\theta_{QCD}\over 32\pi^2} \epsilon^{\mu\nu\alpha\beta}
F^{(a)}_{\mu\nu} F^{(a)}_{\alpha\beta}~.
\label{eq:cpnot}
\end{equation}
Expressing the operator $F\tilde{F}$ in terms of QCD fields, we find the resulting interaction
is proportional to the dot product of the chromoelectric and chromomagnetic fields.
As such, such a term violates $CP$ symmetry and would contribute to the 
electric dipole moment of the neutron. Experimental constraints then 
imply $\theta_{QCD} \le {\cal O}(10^{-9})$. It might be tempting to conclude that $\theta_{QCD}$
is simply absent -- however we know that $CP$ {\it is} violated in the electroweak sector,
and there is no good reason for $\theta_{QCD}$ to be small. This puzzle is known as the
strong $CP$ problem and it is so far unresolved.

\section{Summary and Applications}

\begin{figure*}[t]
\centering
\includegraphics[width=135mm]{qcdsymmetry.epsf}
\caption{A summary of the classical symmetries of QCD and
their quantum fate.} \label{qcdsymmetry}
\end{figure*}

The theme of this talk has been that the origin of mass in QCD is intimately tied to
the classical symmetries of QCD and their quantum fate, summarized  in 
fig. \ref{qcdsymmetry}. The approximate scale symmetry of QCD
is anomalous, giving rise to dimensional transmutation and the scale $\Lambda_{QCD}$.
The dynamical spontaneous breaking of chiral symmetry explains why the pions are light,
while the other baryons and mesons remain heavy. Finally, the $U(1)_A$ anomaly and 
instantons explain why the $\eta'$ is not an approximate Goldstone boson. The
properties of QCD described here have a number of important applications, and we
conclude by mentioning a few of these.

\subsection{Asymptotic Freedom and the Unification of Gauge Couplings}

\begin{figure*}[t]
\centering
\includegraphics[width=60mm]{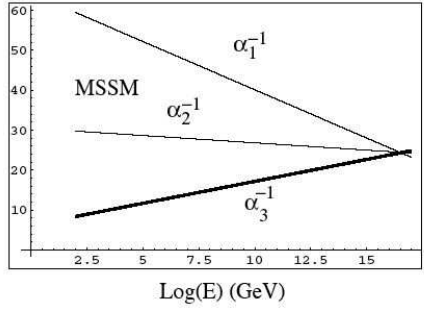}
\caption{An illustration of the unification of couplings
in a supersymmetric model -- asymptotic freedom implies that $\alpha^{-1}_s =
\alpha^{-1}_3$ grows at higher energies.} \label{unify}
\end{figure*}

In Grand Unified Theories \cite{Georgi:1974sy}, one envisions that
all gauge interactions arise from a single gauge theory. In order for this
to occur, all of the gauge-coupling constants must be related. At first glance,
this would seem impossible as their couplings are so different. Asymptotic
freedom, however, implies that the strong coupling becomes smaller at higher
energies -- and at sufficiently high energies, its value can equal \cite{Georgi:1974yf}
that of the
weak or hypercharge couplings.\footnote{Because $U(1)_Y$ is an abelian group, its
normalization is only specified once the unified gauge group $G$ and the embedding of
$U(1)_Y$ in $G$ is specified.} An illustration of the running of the coupling
constants in a supersymmetric model is given in fig. \ref{unify}, and it is the
running of the couplings that determines the scale of the breaking of the unified
gauge group, $M_{GUT} \simeq {\cal O}(10^{16}$ GeV).

\subsection{Top-Quark Matters}

As shown by eqn. (\ref{eq:betacoefficients}), the rate at which the strong-coupling
runs depends on the number of active quark flavors. If one fixes the value of $g_s$
at high energies -- say at the GUT scale -- and then varies the value of the mass of
one of the heavy quarks, one changes the value of $g_s$ at low energies. Hence,
by changing the top-quark mass for fixed high-energy strong coupling, 
one changes the value of $\Lambda_{QCD}$ and hence the mass of the proton.
Using the renormalization group equation, one finds the following dependence
of the proton mass on the top-quark mass \cite{Quigg:1996ew}
\begin{equation}
{m_p \over {1\,{\rm GeV}}} \propto \left({m_t \over 175\,{\rm GeV}}\right)^{2/27}~.
\end{equation}

\subsection{Technicolor}

It is intriguing that the global symmetry breaking structure of two-flavor
QCD, $SU(2)_L \times SU(2)_R \to SU(2)_V$, is precisely the global symmetry structure
of the Higgs sector of the one-doublet standard model. This implies one can
construct a theory of {\it dynamical} electroweak symmetry breaking using
QCD-like dynamics \cite{Weinberg:1975gm,Susskind:1978ms} 
-- these are ``technicolor'' theories \cite{Hill:2002ap}.

In the simplest such model one introduces a new strong $SU(N_{TC})$
gauge theory and, analogous to the up- and down-quarks in QCD, 
two new fermions transforming (which we will
denote $U$ and $D$) as fundamentals
of this gauge symmetry. These new ``techniquarks'' carry an $SU(2)_L \times SU(2)_R$
global symmetry -- the analog of the (approximate) chiral symmetry of the light 
quarks in QCD. Just as in QCD, the ``low-energy'' strong dynamics of this
new gauge theory is expected to cause chiral symmetry breaking, that is a non-perturbative
expectation value for the chiral condensates $\langle \bar{U}_L U_R \rangle =
\langle \bar{U}_R U_L\rangle$ and similarly for the $D$ fermions.

If the left-handed techniquarks form an $SU(2)_W$ doublet, while the right-handed
techniquarks are weak singlets carrying hypercharge, 
technicolor chiral symmetry breaking will result in electroweak 
symmetry breaking. The Goldstone bosons arising from chiral symmetry breaking are
transmuted, by the Higgs mechanism, into the longitudinal components of the electroweak
gauge bosons. 

Theoretically, technicolor addresses all of the shortcomings of the one-doublet
Higgs model: there are no scalars, electroweak symmetry breaking arises in a 
natural manner due to the strong dynamics of a non-abelian gauge theory, 
the weak scale is related to the renormalization group flow of the strong technicolor coupling 
and can be much smaller than any high energy scale and, due to asymptotic freedom, the
theory (most likely) exists in a rigorous sense.

Unfortunately, the simplest versions of this theory -- based, as described, on a scaled-up
version of QCD -- are not compatible with precision electroweak data\footnote{See Langacker and
Erler in \protect\cite{Eidelman:2004wy}.} (and, as described
so far, cannot accommodate the masses of the quarks and leptons). Nonetheless, this simplest
version remains a paradigm for thinking about theories of dynamical electroweak symmetry
breaking.

\section{Conclusions}

As acknowledged by this years Nobel Prize in Physics to Gross, Politzer, and Wilczek, 
the modern understanding of the strong interactions is a great intellectual success story.
As illustrated here, the pieces of this story are highly nontrivial and
hinge on the various ways in which symmetries can be realized (or not) in
quantum field theory. In this sense, and with apologies  to Gilbert
and Sullivan \cite{gilbert}, QCD is {\it the very model of a modern quantum
field theory}.

\begin{acknowledgments}

I am grateful to Fred Gilman, Michael Peskin, Chris Quigg, and especially
to Bogdan Dobrescu, for comments and suggestions.

\end{acknowledgments}


\end{document}